\documentclass[twoside,a4paper]{article}
\usepackage{comment}
\usepackage{amsmath,graphicx,amssymb,fancyhdr,amsthm,,textcomp} 
\usepackage{soul} 
 
\usepackage[usenames]{xcolor}
\usepackage{pdfpages}
\usepackage{authblk}
\usepackage{hyperref}

\theoremstyle{definition} 
  
\theoremstyle{remark}  
  
\def\beq{\begin{eqnarray}}  
\def\eeq{\end{eqnarray}}  
\def\bsp{\begin{split}}  
\def\esp{\end{split}}

\begin{document}

\title{Curvature Invariants in a Binary Black Hole Merger} 
\author[1,2,3 *]{\textsc{Jeremy M. Peters}}
\author[1]{\textsc{Alan Coley}}
\author[2,3,4]{\textsc{Erik Schnetter}}
\affil[1]{Department of Mathematics and Statistics, Dalhousie University, Halifax, Nova Scotia, B3H 3J5, Canada}
\affil[2]{Perimeter Institute for Theoretical Physics, Waterloo, Ontario, N2L 2Y5, Canada}
\affil[3]{Department of Physics \& Astronomy, University of Waterloo, Waterloo, ON N2L 3G1, Canada}
\affil[4]{Center for Computation \& Technology, Louisiana State University, Baton Rouge, LA 70803, USA}
\affil[*]{Corresponding Author}



\date{\today}  
\maketitle  
\pagestyle{fancy}  
\fancyhead{} 
\fancyhead[EC]{}  
\fancyhead[EL,OR]{\thepage}  
\fancyhead[OC]{}  
\fancyfoot{} 

E-mail: \href{mailto:jeremy.peters@dal.ca}{jeremy.peters@dal.ca},\; \href{mailto:alan.coley@dal.ca}{alan.coley@dal.ca}, \href{mailto:eschnetter@perimeterinstitute.ca}{eschnetter@perimeterinstitute.ca}

\begin{abstract}
We study curvature invariants in a binary black hole merger.  It has been conjectured that one could define a quasi-local and foliation independent black hole horizon by finding the level--$0$ set of a suitable curvature invariant of the Riemann tensor.  The conjecture is the geometric horizon conjecture and the associated horizon is the geometric horizon.  We study this conjecture by tracing the level--$0$ set of the complex scalar polynomial invariant, $\mathcal{D}$, through a quasi-circular binary black hole merger.  We approximate these level--$0$ sets of $\mathcal{D}$ with level--$\varepsilon$ sets of $|\mathcal{D}|$ for small $\varepsilon$.  We locate the local minima of $|\mathcal{D}|$ and find that the positions of these local minima correspond closely to the level--$\varepsilon$ sets of $|\mathcal{D}|$ and we also compare with the level--$0$ sets of $\text{Re}(\mathcal{D})$.  The analysis provides evidence that the level--$\varepsilon$ sets track a unique geometric horizon.  By studying the behaviour of the zero sets of $\text{Re}(\mathcal{D})$ and $\text{Im}(\mathcal{D})$ and also by studying the MOTSs and apparent horizons of the initial black holes, we observe that the level--$\varepsilon$ set that best approximates the geometric horizon is given by $\varepsilon = 10^{-3}$.
\end{abstract} 


\section{Introduction}

\subsection{Black Hole Horizons}

Black holes are solutions of general relativity and are most naturally characterized by their event horizon.  The event horizon of a black hole (BH) is defined as the boundary of the causal past of future null infinity.  Intuitively, this means that on the inner side of the event horizon, light cannot escape to null infinity.  Notice that event horizons require knowledge of the global structure of spacetime \cite{AshtekarKrishnan,AshtekarKrishnan2,AshtekarKrishnan3}.  However, for numerical relativity it is more convenient to use an initial value formulation of GR (a 3+1 approach), where initial data is given on a Cauchy hypersurface and is then evolved forward in time.  This approach requires an alternative description of BH horizons which is not dependent of the BH's future. \cite{T,T2,AnderssonMarsSimon,AnderssonMarsSimon2,Jaramillo,Jaramillo2,Hayward:1993,Hayward:1997,Hayward:2005}.  

Let $\mathcal{S}$ be a compact 2D surface without border and of spherical topology, and consider light rays leaving and entering $\mathcal{S}$, with directions $l$ and $n$, respectively.  Let $q_{ab}$ be the induced metric on $\mathcal{S}$ and denote the respective expansions as $\Theta_{(l)} = q^{ab}\nabla_al_b$ and $\Theta_{(n)} = q^{ab}\nabla_an_b$ \cite{news}.  Then, $\Theta_{(l)}$ and $\Theta_{(n)}$ are quantities which are positive if the light rays locally diverge, and negative if the light rays locally converge, and are zero if the light rays are locally parallel.  We say that $\mathcal{S}$ is a {\it trapping surface} if $\Theta_{(l)}<0$ and $\Theta_{(n)}<0$ \cite{AshtekarKrishnan,Penrose:1964wq,ERIK,Schnetter2020}.  Define $\mathcal{S}$ to be a marginally outer trapped surface (MOTS) if it has zero expansion for the outgoing light rays, $\Theta_{(l)} = 0$ \cite{ERIK,Schnetter2020,Schnetter2020_2,Schnetter:2006yt,Schnetter:2006yt2,Schnetter:2006yt3}.  ($\mathcal{S}$ is a future MOTS if $\Theta_{(l)} = 0$ and $\Theta_{(n)} < 0$ and a past MOTS if $\Theta_{(l)} = 0$ and $\Theta_{(n)} > 0$ \cite{Schnetter2020_2}).  MOTSs turn out to be well-behaved numerically, and can be used to trace physical quantities of a BH as they evolve over time and through a BBH merger \cite{Schnetter:2006yt,Schnetter:2006yt2,Schnetter:2006yt3,ERIK,ERIK2}.  

In practice, it is common to view MOTSs as contained in a given 3D Cauchy surface.  Within such a surface, the outermost MOTS is called the apparent horizon (AH) \cite{ERIK,Schnetter2020,Schnetter2020_2,Schnetter:2006yt,Schnetter:2006yt2,Schnetter:2006yt3}.  AHs have many applications to numerical relativity, since tracking an AH only requires knowledge of the intrinsic metric $q_{ab}$ restricted to the spacetime hypersurface and the extrinsic curvature of that hypersurface at a given time \cite{Evans,AshtekarKrishnan3,Schnetter:2006yt2}.  For example, AHs are useful to study gravitational waves, as gravitational fields at the AH are correlated with gravitational wave signals \cite{Evans,6,7,8,Schnetter:2006yt3,news,Iozzo:2021}.  AHs are also used to numerically simulate binary black hole (BBH) mergers and the collapse of a star to form a BH \cite{Booth2005}.  As another example, AHs play a role in checking initial parameters and reading off final parameters of Kerr black holes in gravitational wave simulations at LIGO \cite{Booth2005,LIGO,LIGO2}.  One possible disadvantage of AHs is that the definition of AHs as the "outermost MOTS" relies on the given foliation of the spacetime into Cauchy surfaces \cite{AG2005, Schnetter:2006yt}.   

If one smoothly evolves a given MOTS forward in time, one obtains a world tube which is foliated by these MOTS.  This world tube is known as a dynamical horizon (DH) \cite{BoothFairhurst,AshtekarKrishnan,AshtekarKrishnan2,AshtekarKrishnan3}.  One application of DHs is that they could contribute to our understanding of BH formation \cite{AshtekarKrishnan,AshtekarKrishnan2,AshtekarKrishnan3, Booth2005}.  As is the case with AHs, DHs are dependent on the foliation of the spacetime into Cauchy surfaces, as this spacetime foliation corresponds uniquely to a DHs with a unique foliation into MOTSs \cite{Jaramillo,AG2005,AnderssonMarsSimon,AnderssonMarsSimon2}.  Furthermore, the above definitions of MOTSs, AHs and DHs serve as a quasi-local description of BHs \cite{CMS,CMS2}.

It has been conjectured that one can uniquely define a smooth, locally determined and foliation invariant horizon based on the algebraic (Petrov) classification of the Weyl tensor \cite{CMS, CMS2}.  The necessary conditions for the Weyl tensor to be of a certain Petrov type can be stated in terms of scalar polynomials in the Riemann tensor and its contractions which are called scalar polynomial (curvature) invariants (SPIs).  The first aim of this work is to study certain SPIs numerically during a BBH merger.  

The Petrov classification is an eigenvalue classification of the Weyl tensor, valid in 4 dimensions (D). Based on this classification, there are six different Petrov types for the Weyl tensor in 4D: types ${\bf I}$, ${\bf II}$, ${\bf D}$, ${\bf III}$, ${\bf N}$ and ${\bf O}$ (which is flat spacetime).  One can also use the boost weight decomposition to classify the Weyl tensor, which is equivalent in 4D to the Petrov classification.  One can also algebraically classify the symmetric trace free operator, $S_{ab}$, that is the trace free Ricci tensor, which is equivalent to the Segre classification \cite{kramer}.  

The boost weight algebraic classification generalizes the Petrov classification to $N$-dimensional spacetimes \cite{CMS,CMS2,class,class2,class3,Milson}.  In $N$ D, and with Lorentzian signature $(+1, -1, \ldots, -1)$, we start with the frame of $N$--vectors, $\{{\bf l},{\bf n},\{{\bf m}_i\}_{i = 2}^{N-1}\}$, where ${\bf l}$ and ${\bf n}$ are null and future pointing, ${\bf l}\cdot {\bf n} = 1$, and the $\{{\bf m}_i\}$ are real, spacelike, mutually orthonormal, and span the orthogonal complement to the plane spanned by ${\bf l} $ and ${\bf n}$.  The possible orthochronous Lorentz transformations are generated by null rotations about ${\bf l}$, null rotations about ${\bf n}$, spins (which involve rotations about ${\bf m}_i$), and boosts \cite{Milson}.  With respect to the given frame, boosts are given by the transformation: 
\begin{align*}
    {\bf l} & \rightarrow \lambda {\bf l} \\
    {\bf n} & \rightarrow \lambda^{-1} {\bf n} \\
    {\bf m_i} & \rightarrow {\bf m_i}
\end{align*}
for all $i\in\{2,\ldots,N-1\}$ and for some $\lambda\in \mathbb{R}\backslash \{0\}$.  (The remaining transformations are given in \cite{class,class2,class3,Milson}.)  It is possible to decompose the Weyl tensor into components organized by boost weight \cite{class,class2,class3}.  

It is of particular interest to know whether a given 4D spacetime is of special algebraic type {\bf II} or {\bf D}.  We can state the necessary conditions as discriminant conditions in terms of simple $SPI$s \cite{CMS,CMS2,BIVECTOR,BIVECTOR2}.  Just as an {\it $SPI$} is a scalar obtained from a polynomial in the Riemann tensor and its contractions \cite{CMS,CMS2}, an {\it $SPI$ of order $k$} is a scalar given as a polynomial in various contractions of the Riemann tensor and its covariant derivatives up to order $k$ \cite{CMS,CMS2}.  It turns out that BH spacetimes are completely characterized by their $SPI$s \cite{CH}.  The necessary discriminant conditions on the 4D Weyl tensor for the spacetime to be of type {\bf  II}/{\bf D} can be stated as two real conditions and are given in \cite{CH}.  

Contracting the 4D (complex) null tetrad, $({\bf l},{\bf n},{\bf m},{\overline {\bf m}})$ where ${\bf m}$ and ${\overline {\bf m}}$ are complex conjugates, with the Weyl tensor, $C_{abcd}$, one may form the complex scalars, ${\bf \Psi_0}, {\bf \Psi_1}, {\bf \Psi_2}, {\bf \Psi_3}, {\bf \Psi_4}$ and, in terms of these scalars, as in the Newman-Penrose (NP) formalism (discussed later), one may define the scalar invariants:
\begin{align}
    I &= {\bf \Psi_0}{\bf \Psi_4} - 4{\bf \Psi_1}{\bf \Psi_3} + 3{\bf \Psi_2}^2 \label{I}\\
    J &= \begin{vmatrix}
            {\bf \Psi_4} & {\bf \Psi_3} & {\bf \Psi_2} \\
            {\bf \Psi_3} & {\bf \Psi_2} & {\bf \Psi_1} \\
            {\bf \Psi_2} & {\bf \Psi_1} & {\bf \Psi_0} 
         \end{vmatrix} \label{J}
\end{align}
It can be shown that the two aforementioned real scalar conditions are equivalent to the real and imaginary parts of the following complex syzygy \cite{kramer}: 
\begin{equation}\label{syzygy}
    \mathcal{D} \equiv I^3-27J^2 = 0
\end{equation} 
Thus for Petrov types ${\bf II}$ and ${\bf D}$, equation \eqref{syzygy} holds everywhere.  It also turns out that for Petrov types ${\bf III}$, ${\bf N}$, and for ${\bf O}$, we have $I = J = 0$, so \eqref{syzygy} is satisfied trivially.

\subsection{The Geometric Horizon Conjecture}

Having discussed the Petrov and boost weight classifications, we now turn to the Geometric Horizon Conjecture (GHC) in which we define the geometric horizon (GH) as the set on which the $SPI$s, defined in \eqref{syzygy}, vanish \cite{CMS,CMS2}.  The level--$0$ sets of these $SPI$s might not form a horizon with nice properties, however, since these $SPI$s could vanish additionally on axes of symmetry or fixed points of isometries \cite{CMS,CMS2}.  We know from \eqref{syzygy} that if the spacetime is algebraically special, then the given complex $SPI$ vanishes.  More precisely, the GHC is given as follows \cite{CMS,CMS2}:

\paragraph{GH Conjecture:}
{\em If a BH spacetime is zeroth-order algebraically general, then on the geometric horizon the spacetime
is algebraically special. We can identify this geometric horizon using scalar curvature invariants.}

\paragraph{Comments:}
Note that when studying the GHC, one would need to ensure that the GH exists and is unique.  If the GHC were true in an algebraically general spacetime, then one could say on this horizon, the Weyl tensor is more algebraically special than its background spacetime and this horizon is at least of type ${\bf II}$.  This horizon is then foliation independent and quasi-local \cite{CMS,CMS2}.  

If the spacetime is algebraically special, one then considers the second part of the GHC, which is analogous to the algebraic GHC above, but involving {\it differential $SPI$s}.  Differential $SPI$s (of order $k\geq 1$) are scalars obtained from polynomials in the Riemann tensor and its covariant derivatives and their contractions.  This second part of the GHC thus states that if a BH spacetime is algebraically special (so that on any GH the BH spacetime is automatically algebraically special), and if the first covariant derivative of the Weyl tensor is algebraically general, then on the GH the covariant derivative of the Weyl tensor is algebraically special, and this GH can also be identified as the level--$0$ set of certain differential $SPI$s \cite{CMS,CMS2}.  In this case, the GH is identified as the set of points on which the covariant derivative of the Weyl tensor, $C_{abcd;e}$ is of type ${\bf II}$ \cite{CMS}.  It follows that one may obtain a clearer picture of the GH by taking the level--$0$ sets of these differential $SPI$s.  

In addition to $SPI$s, Cartan invariants can play a role within the frame approach and they are easier to compute.  For example, Cartan invariants are useful in event horizon detection; indeed, it was demonstrated in \cite{GANG,GANG2} that in 4D and 5D, one can construct invariants in terms of the Cartan invariants which detect the event horizon of any stationary asymptotically flat BH solutions.  One could rewrite the statement of the algebraic GHC in the language of the boost-weight classification \cite{Milson} to say that "if there is some frame with respect to which the Weyl tensor in a BH spacetime has a vanishing boost-weight $+2$ term, then on the GH, there is some frame with respect to which the Weyl tensor has a vanishing boost-weight $+1$ term." 
This desired frame is called the {\it algebraically preferred null frame} (APNF).  For example, in an algebraically general 4D spacetime, the APNF is the frame in which the Weyl tensor is of algebraic type {\bf I} so that the boost weight $+2$ terms of the Weyl tensor are $0$ with respect to this frame, which is always possible in 4D.  Then, the GH is identified as the set of points on which the terms of boost weight $+1$  are zero.  (If the Weyl tensor is type ${\bf II}$, then one can analyze the covariant derivative of the Weyl tensor and ask for it to be algebraically special).  The task in this frame approach to study the GHC, therefore, is to first find this APNF, $(l,n,m,\overline{m})$ and thus the AHs/DHs which are orthogonal to $l,n$ \cite{JMMS,AG2005}.  To this end, the Cartan algorithm can be used to completely fix this frame \cite{GANG}, and with respect to this frame, one obtains the associated Cartan scalars.  From these scalars, one can identify the level--$0$ set of $C_{abcd;e}$ with the GH and, via NP calculus, obtain the NP spin expansion coefficients with respect to this APNF.  It is of particular interest to study the spin coefficients, $\rho$ and $\mu$, as their level--$0$ set could be associated with the GH.  However, a more careful study of $\rho$ and $\mu$ and their evolution through a BBH merger is beyond the scope of this paper.  

In this paper, we shall study the (algebraic) $SPI$s in relation to the first (algebraic) part of the GHC.  More specifically, we will study the complex level--zero set of the invariant, $\mathcal{D} = I^3-27 J^2$, as given in \eqref{syzygy}, in 4D during a BBH merger.  This could possibly help provide insight as to whether one can define a proper unique horizon based on the algebraic classification of the Weyl tensor.  This conjecture might have to be modified so that instead of analyzing level--$0$ sets of the real $SPI$s, we analyze instead level--$\varepsilon$ sets for small $\varepsilon$.  Such an $\varepsilon$ could be determined by locating the local minima of the $SPI$s.  However, further evidence from the analysis of $\mathcal{D}_r$ below perhaps suggests that this is not the case.

\subsection{Examples and Motivation}

There are many examples of spacetimes that support the plausibility of the GHC either by explicitly exhibiting GHs or by finding other established BH horizons on which the Weyl and Ricci tensor are algebraically special  \cite{GANG,GANG2,AshtekarKrishnan,AshtekarKrishnan2,AshtekarKrishnan3,CMS,CMS2,szek,kramer}.    For example, in the Kerr spacetime, by invoking the notion of a non-expanding weakly isolated null horizon and an isolated horizon, it can be proven, using the induced metric and induced covariant derivatives on the submanifold and assuming the dominant energy condition, that the Weyl and Ricci tensors are both of type ${\bf II/ D}$ on the event horizon.  This means that one can extract a subset of the set of points where the Weyl and Ricci tensors are both of algebraic type ${\bf II/D}$, to define a smooth BH horizon, namely the event horizon \cite{Ashtekar,Ashtekar2,AshtekarKrishnan,AshtekarKrishnan2,AshtekarKrishnan3,CMS,CMS2}. It can also be shown that the covariant derivatives of the Riemann tensor are of type ${\bf II}$ on this horizon \cite{CMS,CMS2}.  The Kerr geometry approximates the spacetime outside the event horizon of a BH formed by a collapsing star.  By continuity, the region just inside the event horizon must closely approximate the Kerr geometry and it is suggested in \cite{CMS} that there is a surface inside this horizon which is smooth and uniquely identified by geometric constraints \cite{AshtekarKrishnan,AshtekarKrishnan2,AshtekarKrishnan3,Ashtekar,Ashtekar2,CMS,CMS2}, thereby qualifying as a GH. 

Another example to support the GHC comes from a family of exact closed universe solutions to the Einstein-Maxwell equations with a cosmological constant representing an arbitrary number of BHs, discovered by Kastor and Traschen (KT) \cite{KT}.  Consider the merger of 2 BHs.  At early times, there are two 3D disjoint GHs forming around each BH \cite{CMS,CMS2}.  However, at intermediate times, it turns out that the invariant, $\mathcal{D} = I^3 - 27J^2 = 0$ as in \eqref{syzygy} only at the coordinate positions of each of the BHs, along certain segments of the symmetry axis, and along a 2D cylindrical surface, which expands to engulf the 2 BHs as they coalesce \cite{CMS,CMS2}.  During the intermediate process, there are 3D surfaces located at a finite distance from the axis of symmetry for which the traceless Ricci tensor (and hence the Ricci tensor, $R_{ab}$) is of algebraic type ${\bf II/ D}$.  There is also evidence of a minimal 3D dynamically evolving surface where a scalar invariant, ${\cal W}_1$, assumes a constant minimum value.  This suggests that there is a GH during the dynamical regime between the spacetimes \cite{CMS,CMS2}, but further investigation is needed.  At late times, the spacetime then settles down to a type ${\bf D}$ Reissner-Nordstrom-de-Sitter BH spacetime with mass $M = m_1 + m_2$, which turns out to have a GH \cite{GANG,GANG2}.  So a GH forms at the beginning and end of the coalescence.  For further information on the two-BH solution, see \cite{KT}.  The KT solution for multiple BHs was studied and GHs around each BH were found in \cite{CMKT}.  The results were compared with the previously mentioned 2-BH solution.  Additionally, three black-hole solutions were studied and GHs were found around these BHs also \cite{CMS,CMS2}.  For information on more than two BHs, see \cite{KTnsh}.

There are additional examples of spacetimes that support the GHC by identifying GHs with the level--$0$ sets of $\rho$ and $\mu$.  These examples include stationary spacetimes with stationary horizons (e.g. Kerr-Newman-NUT-AdS) \cite{GANG}; spherically symmetric spacetimes such as vacuum solutions or known exact solutions (e.g. Vaidya or LTB dust models) \cite{CMS}; quasi-spherical Szekeres spacetimes \cite{szek}; and the Kastor--Traschen solution for $N>1$ BHs, as mentioned previously \cite{KT}.  The authors have also studied vacuum solutions in the case of axisymmetry (so $R_{ab} = 0$) and where the Weyl tensor is of algebraic type ${\bf I}$ \cite{jpeters:2021}.  Based on the previous examples, it is also natural to study the covariant derivative of the Weyl tensor in this setting and the level--$0$ sets of $\rho$ and $\mu$.

\section{Simulating a Binary Black Hole Merger}

\subsection{Previous Work}

We wish to study the behaviour of the complex $SPI$, $\mathcal{D}$, as defined in \eqref{syzygy}, through a BBH merger.  Since the Kerr geometry is type ${\bf D}$ everywhere, it follows that $\mathcal{D} = 0$ everywhere for a Kerr BH.  Based on our understanding of the features of gravitational collapse and by a plethora of numerical simulations, it is believed that in a BBH merger the merged BHs at late times settle down to a solution well described by the Kerr metric \cite{CMS,CMS2}.  Thus, for a merger of 2 initially Kerr BHs, a plot of the real part and imaginary part of $\mathcal{D}$ should be roughly zero everywhere at early and late times.  However, in the intermediate ``dynamical" region (during the actual merger and coalescence at intermediate times), these same zero plots should yield important information.  This is what we wish to study.  

We highlight some known features of a binary black hole merger, as described in \cite{ERIK,ERIK2,Evans}.  This also serves to set up our notation.  In \cite{ERIK,ERIK2}, it was found that there is a connected sequence of MOTSs, which interpolate between the initial and final states of the merger (two separate BHs to one BH, respectively) \cite{ERIK,ERIK2}.  The dynamics are as follows:  Initially, there are two BHs with disjoint MOTS (which are AHs at this point \cite{Evans}), $\mathcal{S}_1$, and $\mathcal{S}_2$, one around each BH.  Then, as the two BHs evolve, a common MOTS forms around the two separate BHs and bifurcates into an inner MOTS, $\mathcal{S}_i$, which surrounds the MOTS and an outer MOTS, $\mathcal{S}_c$.  $\mathcal{S}_c$ increases in area and encloses $\mathcal{S}_1$, $\mathcal{S}_2$ and $\mathcal{S}_i$, and is the AH at the time of the merger \cite{Evans,ERIK,ERIK2}.  The fate of $\mathcal{S}_c$ and the bifurcation at the time of the merger is well understood \cite{Evans,AshtekarKrishnan2,Schnetter:2006yt,4,Schnetter:2006yt3,ERIK,ERIK2,12,13,Pook-Kolb}.

\begin{figure}
    \centering
    \includegraphics[width = 15cm]{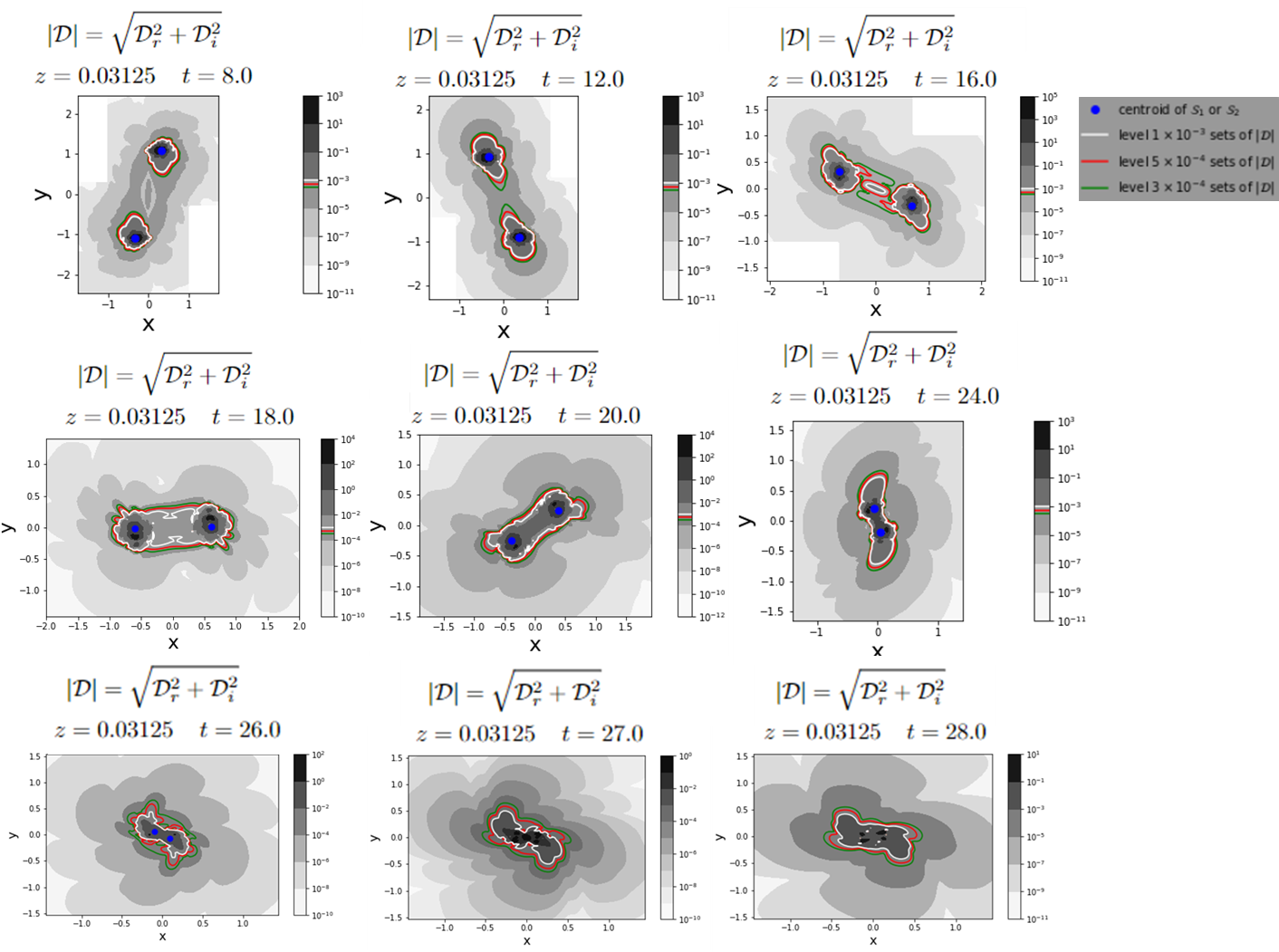}
    \caption{Contour plots of $|\mathcal{D}|$ during a quasi-circular BBH merger consisting of two merging, equal mass and non-spinning BHs at selected times $t = 8$ (upper left), $t = 12$ (upper center), $t = 16$ (upper right), $t = 18$ (middle left), $t = 20$ (center), $t = 24$ (middle right), $t = 26$ (lower left), $t = 27$ (lower center) and $t = 28$ (lower right).}
    \label{1}
\end{figure}

\begin{figure}
    \centering
    \includegraphics[width = 12cm]{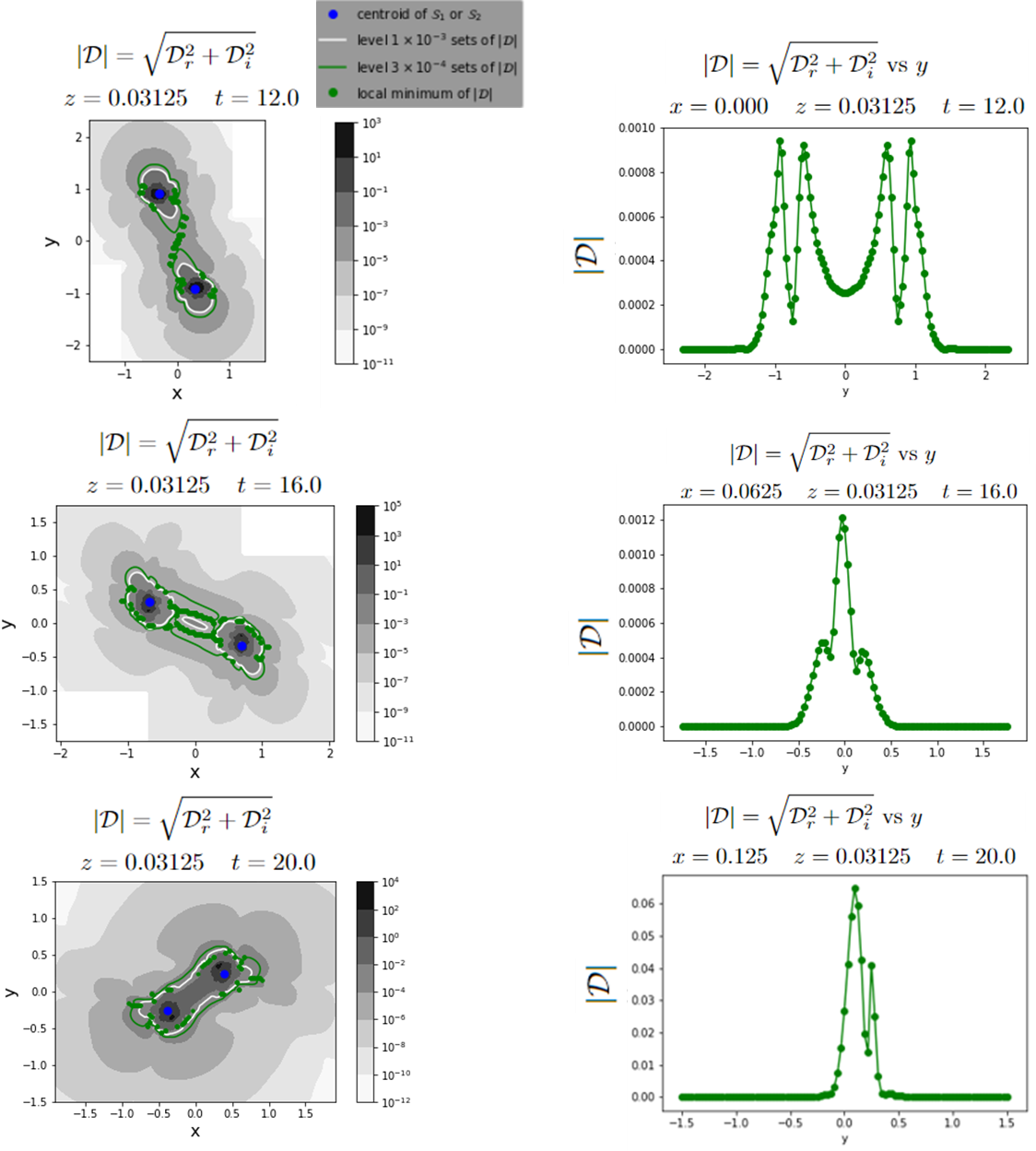}
    \caption{Comparing selected local minima of $|\mathcal{D}|$ along the x–coordinate direction with selected level sets of $|\mathcal{D}|$ at times $t = 12$ (upper left), $t = 16$ (middle left) and $t = 20$ (lower left).  The upper right, middle right, and lower right plots show plots of $|\mathcal{D}|$ vs $y$ for $x = 0.000$ and $t = 12$ (upper right); $x = 0.0625$ and $t = 16$ (middle right); and $x = 0.125$ and $t = 20$ (lower right).}
    \label{2}
\end{figure}

\begin{figure}
    \centering
    \includegraphics[width = 12cm]{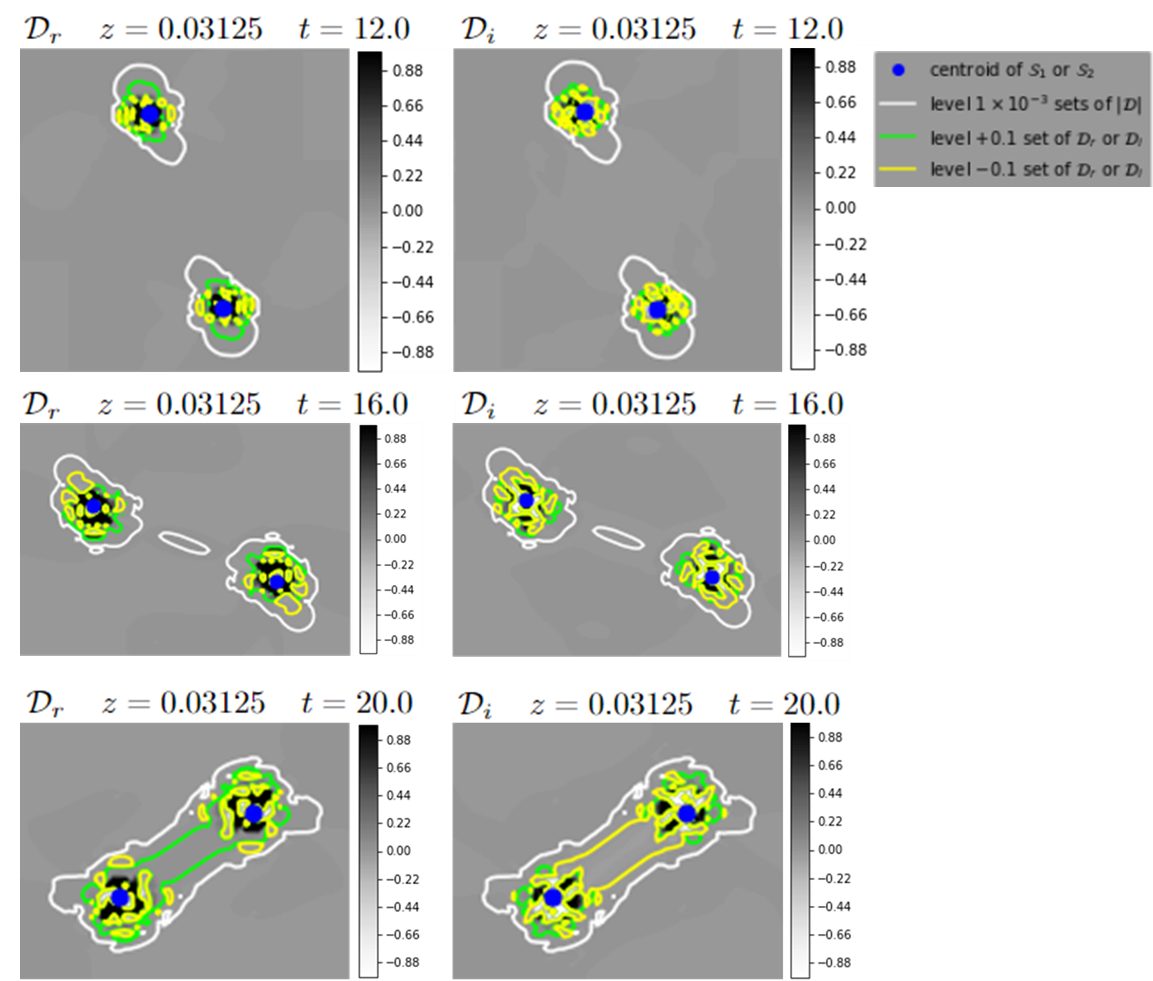}
    \caption{Comparing level–±0.01 contours of $\mathcal{D}_r = \text{Re}(\mathcal{D})$ and $\mathcal{D}_i = \text{Im}(\mathcal{D})$ with level–-$0.001$ contours of $|\mathcal{D}|$.  The upper, middle and lower left plots are plots of $\mathcal{D}_r=\text{Re}(\mathcal{D})$ at times $t = 12,\;16,\;20$, respectively and the upper, middle and lower right plots are plots of $\mathcal{D}_i = \text{Im}(\mathcal{D})$ at times $t = 12,\;16,\;20$, respectively.}
    \label{3}
\end{figure}

\begin{figure}
    \centering
    \includegraphics[width = 12cm]{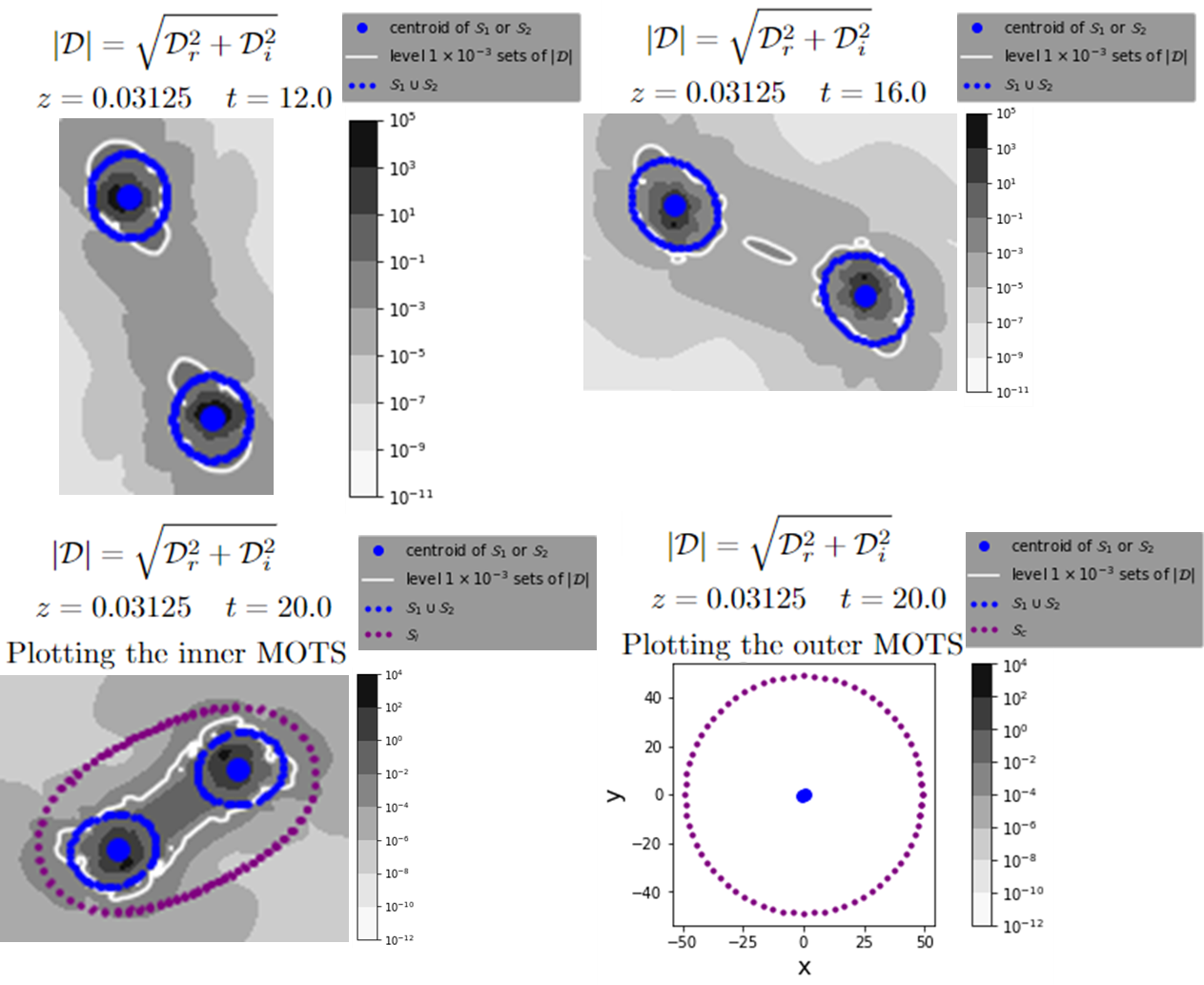}
    \caption{Comparing the white level--$0.001$ sets of $|\mathcal{D}|$ with the MOTS as described in \cite{ERIK,ERIK2} at times $t = 12$ (upper left), $t = 16$ (upper right) and $t = 20$ (lower left and right).  The lower left panel shows the inner MOTS in purple whereas the lower right panel shows the outer MOTS in purple.}
    \label{4}
\end{figure}

\subsection{Present Work}

Instead of studying a head-on collision, in this paper we shall study a quasi-circular orbit of two merging, equal mass and non-spinning BHs.  Apart from \cite{Thesispaper,jpeters:2021}, the analysis of the quantity $\cal{D}$ employed in this simulation is new.  In these simulations, the Einstein toolkit infrastructure was used \cite{Loffler:2011ay} and the simulations are run using $4^{\text{th}}$ order finite differencing on an adaptive mesh grid, with adaptive refinement level of 6 \cite{AMR,AMR2}.  Brill-Lindquist initial data are used, with BH positions and momenta set up to satisfy the initial conditions necessary for a quasi-circular orbit which evolves for less than $1$ orbit before merging (QC0-initial condition).  For more details, see Table I of \cite{QC0}.  Instead of analyzing a sequence of MOTS throughout the merger, we seek to define and study a GH as it evolves through the merger, in accordance with the GHC.  Since \eqref{syzygy} sets necessary conditions for the Weyl tensor to be of algebraic type {\bf II}, we seek to analyze the constant contours of the difference $\mathcal{D} = I^3-27J^2$.  In the simulations, the real and imaginary parts of $I$ and $J$ are calculated using the Weyl scalars, $\{{\bf \Psi}_i\}_{i=0}^5$, as given in equations \eqref{I} and \eqref{J}, and the calculations are carried out using the orthonormal fiducial tetrad, as given by \cite{Lazarus}.  Note that in the rest of this paper, we will use the notation from \cite{ERIK,ERIK2,Evans} to describe the various MOTSs that appear in our simulations.  To recapitulate, $\mathcal{S}_1$ and $\mathcal{S}_2$ are the 1st and 2nd initial MOTS and $\mathcal{S}_i$ and $\mathcal{S}_c$ describe the inner and common MOTS as they appear in our simulation, respectively.  

Figures 1--4 provide plots of various level sets of $\mathcal{D}_r = \text{Re}\{D\}$, $\mathcal{D}_i = \text{Im}\{D\}$ and $|\mathcal{D}|$ as functions of $(x,y)\in \mathbb{R}^2$ at selected instances of the time parameter, $t$, where $t = 0$ indicates the start of the numerical computation.  The level--$\varepsilon$ sets of $|\mathcal{D}|,\mathcal{D}_r,\mathcal{D}_i$ in Figures 1--4 are calculated with a fixed spatial coordinate value of $z = 0.03125$, as are points displayed in Figure 2.  We choose this value of $z = 0.03125$ to illustrate most clearly the main features of the plots.  However, the MOTSs $\mathcal{S}_1\cup\mathcal{S}_2$ in Figure 4 are calculated with $z$ lying in a range $z\in[0.02,0.04]$ and $\mathcal{S}_i$ and $\mathcal{S}_c$ are calculated with $z\in[-0.1,+0.1]$ to accomodate the grid spacing in the spatial coordinates $(x,y,z)$ so that these MOTSs can be displayed fully.  In Figure 4, these level sets of $|\mathcal{D}|,\mathcal{D}_r,\mathcal{D}_i$ are also compared with $\mathcal{S}_1$ and $\mathcal{S}_2$.  The full compliment of pictures describing this BBH merger are displayed in \cite{Thesispaper}.  We present a subset of those figures to illustrate the essential features.  In each of Figures 1-4, the data corresponding to $x<0$ was obtained by rotating the data corresponding to $x>0$ by $180$ degrees about the $x = y = 0$ axis.  In Figures 1--4, we plot the centroids and outlines of $\mathcal{S}_1$ and $\mathcal{S}_2$ along with $\mathcal{S}_i$ and $\mathcal{S}_c$, when they have formed in Figure 4.  The centroids have been added as a marker to track the locations of the BHs through the merger and provide a reference against which we can compare our candidate GHs, namely the level--$\varepsilon$ sets of $|\mathcal{D}|$. 

\subsection{Discussion}

Figure 1 provides the contour plots of the magnitude of $\mathcal{D} = I^3 - 27 J^2$, denoted $|\mathcal{D}|$, on a log scale (see \eqref{syzygy}) for $t = 8,\;12,\;16,\;18,\;20,\;24,\;26,\;27,\;28$ in the upper left, upper center, upper right, middle left, center, middle right, lower right, lower center and lower right panels, respectively, for fixed $z = 0.03125$.  Since $|\mathcal{D}|$ is positive definite, the level--$0$ sets of $|\mathcal{D}|$ are impossible to find precisely due to discrete resolution and numerical error.  Instead, we highlight the evolution of the level--$\varepsilon$ sets, where $\varepsilon = 3\times 10^{-4},\;5\times 10^{-4},\;1\times 10^{-3}$.  The overlaid green, red and white contours of each frame are the level--$3\times 10^{-4}$, level--$5\times 10^{-4}$ and level--$1\times 10^{-3}$ sets of $|\mathcal{D}|$, respectively.

At early times (i.e., at $t = 8$ and $t = 12$ in the upper left and upper center panels, respectively), each of the level--$\varepsilon$ sets are partitioned into pairs of simple closed curves. 
At $t = 16$ (upper right panel), the red level--$5\times 10^{-4}$ set and the white level--$1\times 10^{-3}$ set of $|\mathcal{D}|$ each form a third simple closed curve between $\mathcal{S}_1$ and $\mathcal{S}_2$, which is centred at the origin.  
At times $t = 18$ (middle left panel), $t = 20$ (center panel) and $t = 24$ (middle right panel), for each respective $\varepsilon = 3\times 10^{-4},\;5\times 10^{-4},\;1\times 10^{-3}$, the multiple simple closed curves partitioning the level--$\varepsilon$ set of $|\mathcal{D}|$ have joined so that each level--$\varepsilon$ curve is now a single simple closed curve.  At times $t = 26,\; 27,\; 28$, the level--$\varepsilon$ sets of $|\mathcal{D}|$ continue to track the merged BHs and are displayed in the lower three panels of Figure 1.  It follows that the level--$\varepsilon$ curves for each $\varepsilon = 3\times 10^{-4},\;5\times 10^{-4},\;1\times 10^{-3}$ at each $t$ form an invariantly defined, foliation invariant horizon that contains each separate BH at early times, and contains the merged BH at late times.     

The evolution of the level--$\varepsilon$ curves through the quasi-circular BBH merger in Figure 1 is reminiscent of the sequence of MOTS that take place during the head-on collision simulation in \cite{ERIK,ERIK2}.  In particular, the bifurcation into $\mathcal{S}_i$ and $\mathcal{S}_c$ described in \cite{ERIK,ERIK2, Evans} can be compared to our present quasi-circular simulations.  This comparison is most striking at $t = 16$, in the upper right panel of Figure 1, when the white level--$1\times 10^{-3}$ and red level--$5\times 10^{-4}$ sets are partitioned into three simple closed curves.  However, our numerical computations are not precise enough to study the details of the bifurcation in \cite{ERIK,ERIK2,Evans}.  At times $t = 24,\;26,\;27,\;28$, in the middle right panel and lower three panels of Figure 1 respectively, it also seems that $\mathcal{S}_1$ and $\mathcal{S}_2$ do not merge fully.  Thus, it is possible that at at late times, the level--$\varepsilon$ sets of $|\mathcal{D}|$ for $\varepsilon = 3\times 10^{-4},\;5\times 10^{-4},\;1\times 10^{-3}$ may track $\mathcal{S}_1$ and $\mathcal{S}_2$, which have been found in \cite{Evans} to overlap but not intersect at late times.  However, our simulations did not run to late enough times to make this clear.    

In any case, Figure 1 provides strong evidence that for each $\varepsilon = 3\times 10^{-4}, \;5\times 10^{-4}, \;1\times 10^{-3}$, the level--$\varepsilon$ sets of $|\mathcal{D}|$ track a unique GH, which can be identified by the level--$0$ set of $\mathcal{D}$.  It is of interest to study the level--$\varepsilon$ curves as $\varepsilon\rightarrow 0$ and extrapolate from our results the appearance of level--$\varepsilon$ curves for arbitrary small $\varepsilon$.  This could be aided with improved numerical resolution.  Such an extrapolation scheme is beyond the scope of this paper, however, and in the meantime we study the features of level--$\varepsilon$ curves for an appropriate value of $\varepsilon$.

We observe that for $\varepsilon = 3\times 10^{-4},\;5\times 10^{-4},\;1\times 10^{-3}$, the level--$\varepsilon$ contours are very close to each other, showing that the level--$\varepsilon$ sets vary continuously with $\varepsilon$.  We also observe that if $\varepsilon_1 \leq \varepsilon_2$, then the 2D area enclosed by the level--$\varepsilon_1$ curve encloses the 2D area enclosed by the level--$\varepsilon_2$ curve.  Thus, each panel of Figure 1 indicates that $|\mathcal{D}|$ decreases on average away from $\mathcal{S}_1$ and $\mathcal{S}_2$, and the plots of $|\mathcal{D}|$ show no global minima.  However, Figure 2 indicates that the plots of $|\mathcal{D}|$ do have have local minima which approximately coincide with the level--$0$ sets of $\mathcal{D}_r$ and with the level--$\varepsilon$ sets for $\varepsilon = 3\times 10^{-4},\;1\times 10^{-3}$.

In order to investigate further the level--$0$ sets of $|\mathcal{D}|$ (or equivalently the level--$0$ sets of $\mathcal{D}$), we find out where $|\mathcal{D}|$ takes a local minimum value.  If the value of $|\mathcal{D}|$ itself is small, then these locations of the local minima could possibly indicate positions of the actual zeros of $|\mathcal{D}|$, which would be caused by numerical errors.  
It could also be the case that the GHC should be modified so that the GH is defined as the set of points where $|\mathcal{D}|$ reaches the local minimum instead of being identically zero.  If this were the case, then locating the local minima of $|\mathcal{D}|$ would locate the GH precisely instead of approximating it.  However, further evidence from the analysis of $\mathcal{D}_r$ below perhaps suggests that this is not the case.  

To this end, in Figure 2, we examine 1D plots of $|\mathcal{D}|$ as functions of $y$ for fixed $x = x_0$.  At time $t = 12$, (resp. $t=\;16,\;20$), we display the $|\mathcal{D}|$ vs $y$ plot in the upper-right panel (resp. middle-right panel, lower-right panel) of Figure 2 for $x_0 = 0.000$ (resp. $x = 0.0625,\;0.125$).  Along each $|\mathcal{D}|$ vs $y$ plot, we find and track the values of $y = y_{min}$, where $|\mathcal{D}|$ assumes a local minimum value and lies in the range $\left[1\times 10^{-4}, 1.2\times 10^{-3}\right]$.  The resulting coordinates $(x_0,y_{min})$ are then superimposed on the level--$1\times 10^{-3}$ and level--$3\times 10^{-4}$ sets of $|\mathcal{D}|$ at $t = 12$ (resp. $t = 16,\;t = 20$) in the upper-left panel (resp. middle-left, lower-left) panel of Figure 2.    

More specifically, in the upper-right panel of Figure 2, where $t = 12$ and $x_0 = 0.000$, we see that local minima of $|\mathcal{D}|$ occur roughly at $y_{min} = -0.8,\;0,\;0.8$.  The points $(x_0,y_{min}) = (0,-0.8), (0,0), (0,+0.8)$ are then marked with green dots on the level--$\varepsilon$ plots of $\mathcal{D}$ at $t = 12$, as shown on the upper-left panel of Figure 2.  The remaining green dots on this upper-left panel are found similarly, using the positions $(x_0,y_{min})$ of the local minima of $|\mathcal{D}|$, with $|\mathcal{D}|\in \left[1\times 10^{-4}, 1.2\times 10^{-3}\right]$, but with varying $x_0$.

One can similarly inspect the plot in the middle-right panel of Figure 2, where $t = 16$ and $x_0 = 0.0625$, to find that $|\mathcal{D}|$ is minimized roughly where $y_{min} = -0.2,\;0.1$.  The corresponding points $(x_0,y_{min}) = (0.125, -0.2), (0.125, 0.1)$ are then recorded with green dots on the level--$\varepsilon$ plots of $\mathcal{D}$ at $t = 16$, in the middle-left panel of Figure 2.  The remaining green dots on this panel are again obtained by varying $x_0$ and mark the positions $(x_0,y_{min})$ of the local minima of $|\mathcal{D}|\in\left[1\times 10^{-4}, 1.2\times 10^{-3}\right]$.

Finally, in the lower-right panel of Figure 2, where $t = 20$ and $x_0 = 0.125$, the quantity $|\mathcal{D}|$ is minimized at roughly $y_{min} = 0.2$, so that the point $(x_0,y_{min}) = (0.125, 0.2)$ is marked with a green dot on the lower-left panel of Figure 2 and the remaining green dots track the positions $(x_0,y_{min})$ of the local min of $|\mathcal{D}|$, $|\mathcal{D}|\in\left[1\times 10^{-4}, 1.2\times 10^{-3}\right]$, as above. 

By studying the left three panels of Figure 2, we observe that at times $t = 12$ and $t = 16$, these local minima appear to track the green level--$3\times 10^{-4}$ sets, while at time $t = 20$, these local minima appear to track more closely the white level--$1\times 10^{-3}$ sets.  This shows that the positions of the local minima of $|\mathcal{D}|$ align closely with the level-$\varepsilon$ sets of $|\mathcal{D}|$ for $\varepsilon = 3\times 10^{-4},\;1\times 10^{-3}$.  In \cite{Thesispaper}, the positions of $(x_0,y_{min})$ are compared with the positions of the minima $(x_{min},y_0)$--obtained through the above procedure but with the roles of $x$ and $y$ reversed--and it is possible that $(x_{min},y_0)$ are indeed the overall local minima of $|\mathcal{D}|$.  In this case, Figure 2 provides supporting evidence that the local minima of $|\mathcal{D}|$ and hence the level--$\varepsilon$ sets of $|\mathcal{D}|$, accurately track the GH.  It remains as future work, however, to study the local minima more closely and devise and implement an algorithm to track its $(x,y)$ position through a BBH merger.

When studying the zero set of $\mathcal{D}$, it is also helpful to study separately $\mathcal{D}_r = \text{Re}(\mathcal{D})$ and $\mathcal{D}_i = \text{Im}(\mathcal{D})$, as these quantities change sign through a zero.  To wit, we plot in Figure 3 the contour plots of $\mathcal{D}_r$ (left panels) and $\mathcal{D}_i$ (right panels), with magnified resolution, along with their level--$-0.01$ sets in yellow and their level--$+0.01$ sets in lime green at times $t = 12,\;16,\;20$ in the upper, middle and lower panels, respectively.  

In each of the frames in Figure 3, the grey regions correspond to $-0.01<\mathcal{D}_r<0.01$ (resp. $-0.01<\mathcal{D}_i<0.01$), the black regions correspond to $\mathcal{D}_r\geq 1$ (resp. $\mathcal{D}_i\geq 1$), and the white regions correspond to $\mathcal{D}_r\leq -1$ (resp. $\mathcal{D}_i\leq -1$).  Interpolating between the level--$+0.01$ sets and the level--$-0.01$ sets of $\mathcal{D}_r$ (resp. $\mathcal{D}_i$), we deduce that there must be a surface among the level--$\pm0.01$ sets of $\mathcal{D}_r$ (resp. $\mathcal{D}_i$) across which $\mathcal{D}_r$ (resp. $\mathcal{D}_i$) changes sign.  This surface is then the level--$0$ set of $\mathcal{D}_r$ (resp. $\mathcal{D}_i$).

Upon inspection of each frame in Figure 3, we see that the level--$\pm0.01$ sets of $\mathcal{D}_r$ (resp. $\mathcal{D}_i$) occur in close proximity with, but are contained in, the interior of the level--$1\times 10^{-3}$ sets of $|\mathcal{D}|$.  Therefore, Figure 3 provides strong evidence that the level--$1\times 10^{-3}$ set of $|\mathcal{D}|$ well approximates the level--$0$ set of $\mathcal{D}_r$ and $\mathcal{D}_i$ and, hence, the level--$0$ set of $\mathcal{D}$.  

We next explicitly compare the level--$1\times 10^{-3}$ sets of $|\mathcal{D}|$ with the corresponding MOTSs $\mathcal{S}_{1,2,i,c}$ in Figure 4.  We display the 2D contour plots of $|\mathcal{D}|$ with magnified resolution at times $t = 12$ and $t = 16$ in the upper left and upper right panels, respectively, and we display the 2D contour plots of $|\mathcal{D}|$ at $t = 20$ and in the lower left and lower right panels.  As in Figures 1--3, the white curves denote the white level--$1\times 10^{-3}$ sets of $|\mathcal{D}|$ and the blue curves mark the $(x,y)$ coordinates of points on $\mathcal{S}_1\cup\mathcal{S}_2$ whose corresponding $z$ coordinate values lie in the range $\left[0.02,0.04\right]$, as mentioned previously.

In the present quasi-circular simulation, the bifurcation of the the third MOTS into $\mathcal{S}_i$ and $\mathcal{S}_c$ occurs between times $t = 18.5$ and $t = 18.75$.  Once this happens, $\mathcal{S}_1$ and $\mathcal{S}_2$ are no longer AHs, as the MOTS, $\mathcal{S}_c$, now surrounds $\mathcal{S}_1$, $\mathcal{S}_2$ and $\mathcal{S}_i$.  Thus, in order to compare our level--$\varepsilon$ sets of $|\mathcal{D}|$ with AHs (the outermost MOTSs), we have included plots of $\mathcal{S}_i$ and $\mathcal{S}_c$ here.  The purple dots on the bottom left (resp. bottom right) panel of Figure 4 label the $(x,y)$ coordinates of the points on $\mathcal{S}_i$ (resp. $\mathcal{S}_c$) whose corresponding $z$ value lies in the range, $\left[-0.1,+0.1\right]$.  Note that in the bottom right--hand corner, the outer MOTS at late times is so big that the entire two dots and the scale for scale of the white level--$0.001$ sets of $|\mathcal{D}|$ and $\mathcal{S}_{i,1,2}$ are squashed at the origin.

We find that the white level--$1\times 10^{-3}$ set of $|\mathcal{D}|$ coincides closely with $\mathcal{S}_1$ and $\mathcal{S}_2$, especially at early times.  Since the MOTSs $\mathcal{S}_c$ and $\mathcal{S}_i$ have not yet formed, $\mathcal{S}_1$ and $\mathcal{S}_2$ are AHs at early times.  Hence, Figure 4 shows that the white level--$1\times 10^{-3}$ sets of $|\mathcal{D}|$ well approximate the AH at early times.  This lends support to the choice of the white level--$1\times 10^{-3}$ sets of $|\mathcal{D}|$ as a representative approximation to the level--$0$ set of $|\mathcal{D}|$ (and hence of $\mathcal{D}$).

At later times, however, it appears that the AH, $\mathcal{S}_c$, diverges from the white level--$1\times 10^{-3}$ sets, so that the white level--$1\times 10^{-3}$ sets of $|\mathcal{D}|$ no longer approximate the AH in this r\'{e}gime.  Thus, it appears in this current simulation that the level-sets of the curvature invariant $\mathcal{D}$ does not detect the common outer horizon forming at late times.  Furthermore, in this simulation, no significant patterns were observed in the level--$\varepsilon$ sets of $|\mathcal{D}|$ immediately prior to the formation of $\mathcal{S}_i$ and $\mathcal{S}_c$ \cite{Thesispaper}.  However, the authors intend to further study the GHs at the formation of $\mathcal{S}_i$ and $\mathcal{S}_c$, and also at later times, which would require rerunning these simulations with improved numerical resolution.          

Therefore, Figures 1--4 provide strong evidence that one can define a unique smooth GH, theoretically given by the level--$0$ set of the complex invariant $\mathcal{D} = I^3 - 27J^2$, which we have found is best approximated in the numerics by the level--$1\times 10^{-3}$ sets of $|\mathcal{D}|$.

\section{Conclusions}

We have studied the algebraic properties of the Weyl tensor by analyzing the time evolution of various level--$\varepsilon$ sets of $|\mathcal{D}|$ through a quasi-circular merger of two non-spinning, equal mass BHs where, in particular, $\varepsilon = 3\times 10^{-4},\;5\times 10^{-4},\;1\times 10^{-3}$.  These level--$\varepsilon$ contours are superimposed on the contour plots of $|\mathcal{D}|$ in Figure 1.  In these plots, the locations of the two initial BHs were tracked by using the centroids of the initial AHs.  We found that at early times, each such level set is partitioned into two disjoint simple closed curves, each of which contains one of the two centroids of the AHs of the 2 separate initial BHs.  Then each level set, at some intermediate time, forms a third simple closed curve which is centred at the origin and positioned between the centroids of the AHs of the two initial BHs.  These three simple closed curves then join and form one simple closed curve for each level set, which contains the centroids of both initial BHs.  

The plots for $|\mathcal{D}|$ in Figure 1 provide strong evidence that the level sets of $|\mathcal{D}|$ identify the GH.  However, it is impossible to identify the level--$0$ sets of $|\mathcal{D}|$ precisely, since $|\mathcal{D}|$ is a sum of positive definite terms, so numerical errors and discrete resolution cause $|\mathcal{D}|$ to be strictly positive.  Thus, to further study the zeros of $|\mathcal{D}|$, which would indicate the zeros of the complex quantity $\mathcal{D}$, we studied the positions of the local minima of $|\mathcal{D}|$ along plots of $|\mathcal{D}|$ vs $y$ for a fixed $x$ in Figure 2.  Figure 2 demonstrates that the level--$\varepsilon$ sets of $|\mathcal{D}|$ correspond closely to the local minima of $|\mathcal{D}|$, where $\varepsilon = 3\times 10^{-4},\;1\times 10^{-3}$.  Since the local minima of $|\mathcal{D}|$ approximate the zeros of $|\mathcal{D}|$, Figure 2 provides supporting evidence that the level--$\varepsilon$ sets of $|\mathcal{D}|$ for $\varepsilon = 3\times 10^{-4},\;1\times 10^{-3}$ track the GH of the BBH merger. 

Since $|\mathcal{D}|$ is positive definite, its zeros cannot be traced by positive and negative level sets.  Therefore, we have also analyzed quantities which change sign through a zero.  In Figure 3, we examined contour plots of $\mathcal{D}_r = \text{Re}(\mathcal{D})$ and $\mathcal{D}_i = \text{Im}(\mathcal{D})$ with their associated level--$\pm0.01$ sets and compared these plots with the white level--$1\times 10^{-3}$ sets of $|\mathcal{D}|$.  We found surfaces surrounding the union of the level--$\pm0.01$ contours of $\mathcal{D}_r$ (resp. $\mathcal{D}_i$) across which $\mathcal{D}_r$ (resp. $\mathcal{D}_i$) change sign, and are hence a subset of the level--$0$ sets of $\mathcal{D}_r$ (resp. $\mathcal{D}_i$).  We also found these particular zeros of $\mathcal{D}_r$ (and those of $\mathcal{D}_i$) to be well approximated by the level--$1\times 10^{-3}$ contours of $|\mathcal{D}|$ in our plots and suggest that this approximation is valid in this setting.

In Figure 4, we compare the level--$1\times 10^{-3}$ contours of $|\mathcal{D}|$ with the AHs.  These AHs are given by $\mathcal{S}_1$ and $\mathcal{S}_2$ at early times and by $\mathcal{S}_c$ after $\mathcal{S}_c$ has formed.  Figure 4 shows that the level--$1\times 10^{-3}$ sets of $|\mathcal{D}|$ are very well approximated by the AHs, $\mathcal{S}_1$ and $\mathcal{S}_2$, at early times, but at later times the AH diverges from this level set of $|\mathcal{D}|$.  However, even at late times, $\mathcal{S}_1$ and $\mathcal{S}_2$ continue to track a subset of the level--$1\times 10^{-3}$ set of $|\mathcal{D}|$, as does $\mathcal{S}_i$.

{\em Therefore, in the binary black hole merger, as displayed in Figures 1--43 in \cite{Thesispaper} and summarized in Figures 1--4 above, the algebraic structure of the Weyl tensor is clearly identified by the level--$\varepsilon$ sets of $|\mathcal{D}|$, and it is plausible that the level set with $\varepsilon = 1\times 10^{-3}$ accurately identifies the geometric horizon}.

However, there is much future work still to be done.  It is of interest to rerun these simulations, but at a higher numerical resolution to analyze more systematically the quantity $|\mathcal{D}|$ and its level sets.  In particular, it remains to study the behaviour of the level--$\varepsilon$ curves of $|\mathcal{D}|$ as $\varepsilon\rightarrow 0$ and compare these level sets with the respective features in Figures 2--4.  It would also be of interest to study in greater detail the local minima of $|\mathcal{D}|$ and track its evolution through the BBH merger.  Thirdly, it remains to study the evolution of the level--$\varepsilon$ curves immediately prior to the formation of $\mathcal{S}_{i,c}$, and at late times, when $\mathcal{S}_c$ is the AH.  Finally, the authors plan to study the time evolution of the covariant derivative of the Weyl tensor through a BBH merger in the context of the APNF approach.

\section{Acknowledgements}  
 
This work was supported financially by NSERC (AAC and ES).  JMP would like to thank AAC for supervising his masters thesis and ES for numerical assistance and useful discussions, and the Perimeter Institute for Theoretical Physics for hospitality during this work.  

\section{Data Availability}

Complete data for this work is presented in \cite{Thesispaper}, which is available on\newline \url{https://dalspace.library.dal.ca/handle/10222/79721},\newline and is also available from the corresponding author on reasonable request.


\bibliographystyle{plain}
\bibliography{sources.bib}

\end{document}